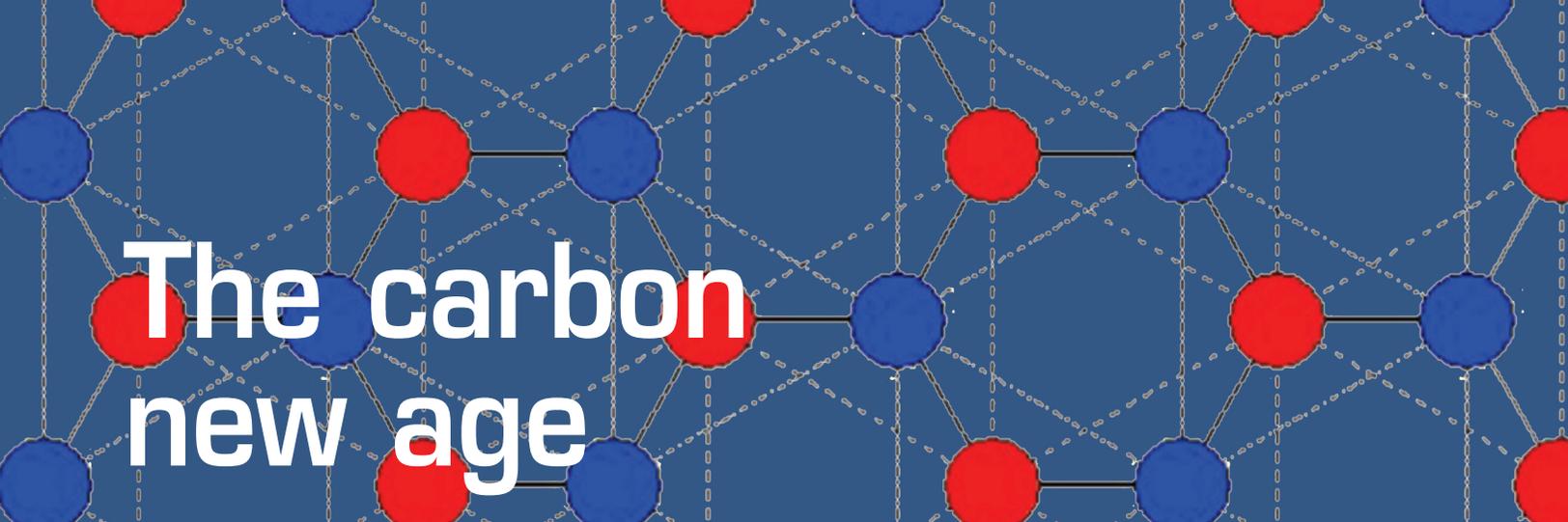

# The carbon new age

Graphene has been considered by many as a revolutionary material with electronic and structural properties that surpass conventional semiconductors and metals. Due to its superlative qualities, graphene is being considered as the reference material for a post-CMOS technology. Furthermore, graphene is also quite unusual electronically since its electric carriers behave as if they were massless and relativistic, the so-called Dirac particles. Because of its exotic electronic properties, theorists are being forced to revisit the conceptual basis for the theory of metals. Hence, graphene seems to be unveiling a new era in science and technology with still unseen consequences.

Antonio H. Castro Neto
*Department of Physics, Boston University, 590 Commonwealth Avenue, Boston, MA 02215 USA.*
*E-mail: neto@bu.edu*

Human progress and development has always been marked by breakthroughs in the control of materials. Since pre-historic times, through the stone, bronze, and iron ages, humans have exploited their environment for materials that can be either used directly or can be modified for their benefit (bronze is an alloy made of cooper and tin), to make their life more comfortable, productive, or to give them military advantage. One age replaces another when the material that is the basis for its sustainability runs its course and is replaced by another material which presents more qualities (for instance, iron is lighter and stronger than bronze).

In the 20$^{th}$ century the advancements in material science have been so profound that the process of material "raise and fall" has accelerated tremendously. This process has shaped in such a strong way our lives that it is very hard to imagine how the world looked like before jet planes, computers, and the global unification that the internet has created. This revolution in the way we interact with the world has its origin in the advances made in the beginning of the 20th century with the advent of quantum mechanics.

Quantum physics and chemistry are the cornerstones for the understanding on how materials behave electronically and structurally, and hence, the basis for essentially everything that surround us. Since quantum mechanics was firmly established theoretically and experimentally, a plethora of new materials, which today make our way of life, have been created: plastics, rubbers, glasses, metallic alloys, semiconductors, superconductors, and magnets, just to mention some.

The so-called "silicon age" dominated the last half of the 20$^{th}$ century and extends to this very day. The domination of control over semiconductors such as silicon, was fundamental for the creation of the globalized world as we know it. Ultra-fast computers and communication are at the heart of our society and have spread over the entire world. Even in the most recondite places of our planet there is a human being either using a computer, a cell phone, or a GPS.

However, just like any other materials human beings have played with over millennia, semiconductors such as silicon and gallium arsenide are reaching their "sunset".

The International Technology Roadmap for Semiconductors[1] (ITRS), which is sponsored by the five leading chip manufacturing regions in the world (Europe, Japan, Korea, Taiwan, and the United States), and has the objective of ensuring cost-effective advancements in the performance of the integrated circuit, has clearly identified an end-of-life for scaled complementary metal–oxide–semiconductor (CMOS) technology around 2022. The causes for the demise of silicon-based technology range from purely physical to economical. Silicon, as a crystalline solid, ceases to exist beyond the 10 nanometers (1 nm = $10^{-9}$ m = 10 Å) because thermal fluctuations make atoms fluctuate strongly as the dimensions of the material are reduced transforming crystals into amorphous material. Moreover, in order to reduce the volume of a microchip by a factor of two requires a factor ten of investment in new technology.

However, in a recent workshop, sponsored by ITRS, on "Beyond CMOS" technologies, the final reports states that "Carbon-based Nanoelectronics has a major advantage in that science and technology resulting from accelerated development in carbon nanotubes (CNT) and graphene nano-ribbons GNRs for metal–oxide–semiconductor field-effect transistor (MOSFET) applications can provide substantial basis for exploring and developing new physical phenomena in these materials "Beyond CMOS" information processing paradigm." There are several reasons why ITRS has identified carbon for the "Beyond CMOS" technologies.

From the history of materials perspective, it seems almost natural to think about carbon as the next platform for micro-electronics. The first transistors developed in the Bell Labs during the 1950's were based on germanium and only later they evolved to silicon. The reasons for the replacement of Ge by Si involve the fact that Si, after enough processing, can be made very pure and hence conduct electricity with much less energy loss, and at a much lower cost (Si costed $10 per kg compared to Ge which was almost at $1 800 per kg - compare with carbon, in graphite form, that costs around $2 per kg !). A quick look at the column IV A of the periodic table of elements (see Fig. 1) shows that evolution from Ge to Si is associated with a jump of one row in the table and that the ultimate material in this column is exactly carbon.

Moving up in the column IV A has many physical and chemical implications that can be understood from basic quantum mechanics and the fact that electrons interact among themselves through Coulomb forces. Although these elements have the same number of electrons in their topmost electronic shell, and hence similar chemical properties, the size of the electronic wave-functions and their energy vary a great deal. For instance, a Si atom has eight more electrons than a C atom. Hence, Si has larger electronic clouds that act to screen or shield the Coulomb interaction between the electrons. This is why the chemical bonds in Si crystals are weaker and longer than in C crystals (2.35 Å in Si and 1.42 Å in C). C has the highest melting and sublimation temperatures of all elements at about 3 500 °C, while Si melts at 1 700 °C. These differences help to understand why C is the element responsible for life while Si only appears in rocks and sand on earth.

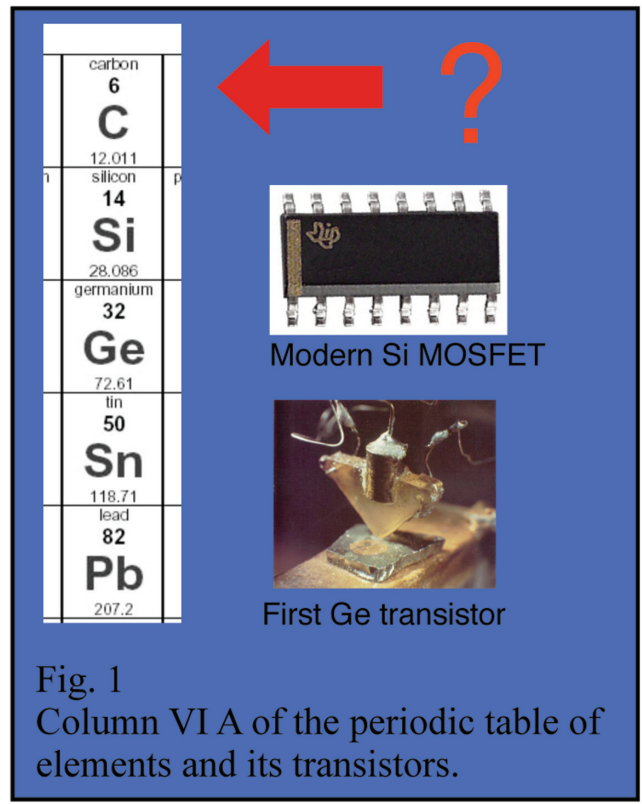

Fig. 1 Column VI A of the periodic table of elements and its transistors.

One of the most interesting aspects of C chemistry is the fact that its electronic states are better described in terms of the hybridization of pure s and p hydrogen-like states. These hybridized orbitals form strong directional covalent bonds leading to a large number of different crystal structures, or allotropes. In Fig. 2 we show some of the most important allotropes of pure carbon. Close inspection of these allotropes show that they all have the same basic motif, namely, the benzene ring. Linus Pauling was one the first scientists to understand the nature of these allotropes and in his 1950's masterpiece "The Nature of the Chemical Bond", Pauling describes graphite (the only allotrope that was known during this time – another example on how fast material science has evolved in the last 50 years) as made out of a layers of a "giant molecule" that we today call graphene. Graphene can be considered the "mother" of all allotropes shown in Fig. 2: graphite is stacked graphene, nanotubes are rolled graphene, and fullerenes are wrapped graphene. In fact, most of the electronic and structural properties of these allotropes can be derived from the basic properties of graphene.

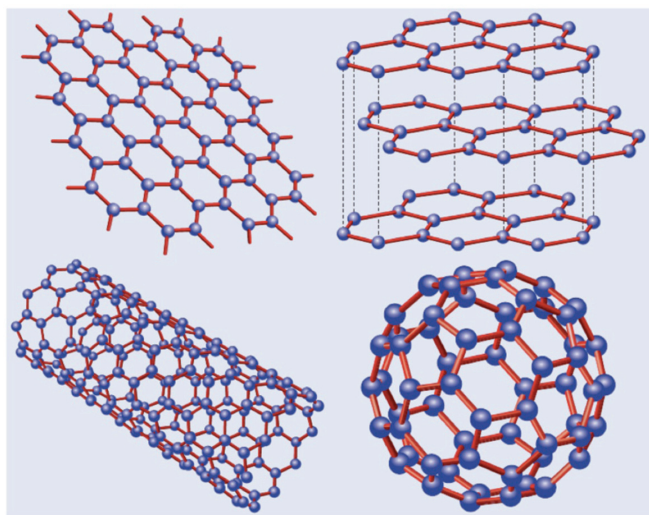

Fig. 2 Carbon allotropes[6]. Top left: Graphene; top right: Graphite; bottom left: nanotube; bottom right: fullerene.

Graphene was the last allotrope to be discovered. While fullerenes were discovered in the 1980's and nanotubes in the 1990's, graphene was only discovery in 2004 by the group lead by Andre Geim at Manchester University in England[2]. Amazingly, graphene was discovered by exfoliation (or peeling) of graphite using what nowadays is called "scotch tape technique".

After its discovery there was a short period of time when graphene was seen either as mere curiosity or even with distrust, since it was believed that such a material could not exist in crystalline form. The field of graphene research only really took over in 2005 with the measurement of the anomalous Hall effect by Geim's[3] and Philip Kim's[4] group at Columbia University in the USA (see Fig. 3). These experiments showed, without any question, that the electrons in this material behaved in a "relativistic" way, although they move at velocities 300 times smaller than the speed of light. By showing that the electrons behave as so-called Dirac particles[5], Geim and Kim have proved that Pauling's "giant molecule" is actually a crystal since Dirac electrons can only exist on the honeycomb lattice (see Fig. 2). The fact that at low energies and long wavelengths the graphene electrons behave as relativistic particles has attracted the attention of the whole condensed matter community[6]. It opened the field to the possibility of performing experiments that were proposed in the realm of particle and nuclear physics but that were never tried before because they required extreme conditions such as enormous electric and magnetic fields, such as the fields that only exist in the vicinity of neutron stars or black-holes.

Although graphene is one atom thick, it can be seen with an ordinary optical microscope when placed on top of a properly chosen $SiO_2$ substrate[1]. It has the properties of a good metal, although its electronic properties do not fit the standard theory of metals[5]. Graphene is also resistant against extrinsic impurities because its chemical bonding is very specific and consequently graphene conducts electricity better, with less energy loss, than any other semiconductor, including Si, and even Cu[7]. Moreover, graphene is one of the strongest materials ever measured in terms of Young's modulus and elastic stiffness[8] (the only other material that is comparable in strength is diamond - another carbon allotrope), nevertheless it is one of softest (the only example of a metallic membrane[9]). It can be used as an ultra-sensitive nano-mechanical resonator besides being highly impermeable[10]. Hence it is not surprising that so many high-tech industries are interested in developing graphene-based devices for a plethora of applications, from high-frequency transistors[11] to surface coating.

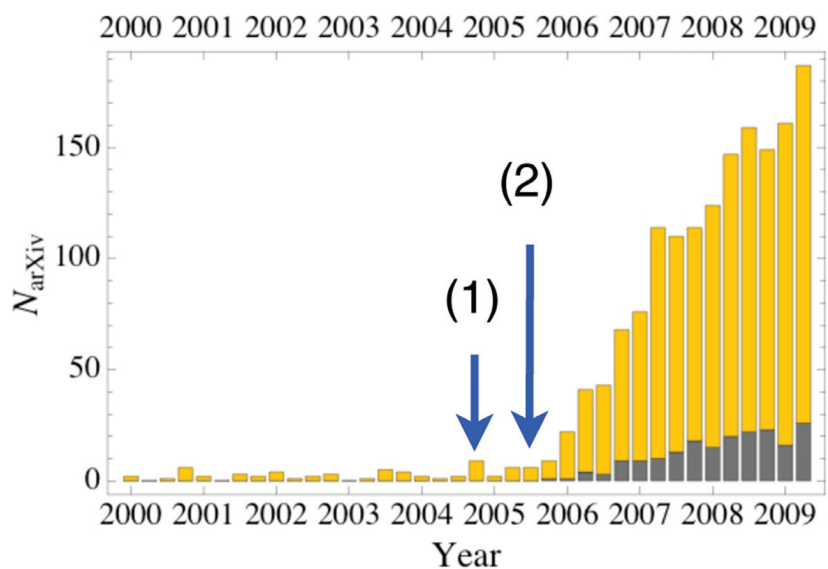

Fig. 3 Number of papers posted quarterly on arxiv.org on graphene (yellow) and bilayer graphene (gray) as a function of time. (1) Discovery of graphene; (2) Anomalous QHE measurement.

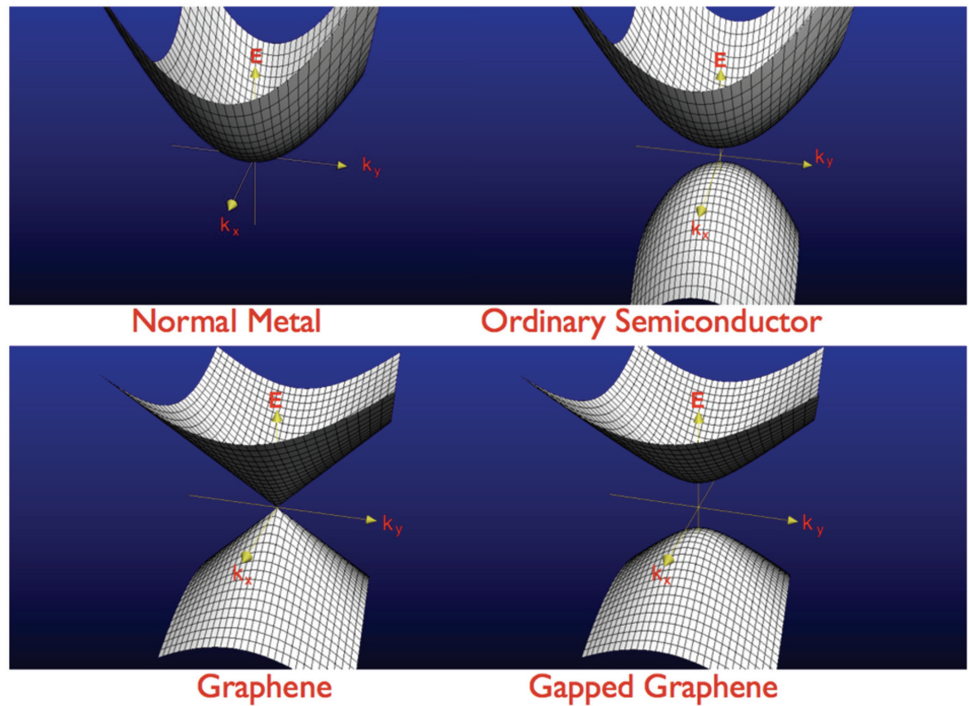

*Fig. 4 Energy-momentum relations for different types of materials.*

In fact, graphene can overcome some of the major barriers that have been found in using other C allotropes, such as nanotubes, for practical applications in transistors, sensors, etc. For one, it does not suffer from the geometric limitations of a one-dimensional material and its properties can be tailored more easily. However, there are still major challenges that have to be surmounted before graphene can play any role in the electronic industry. It is actually unreasonable to think that graphene will replace Si in the electronic industry anytime soon. However, it is fair to imagine that some form of modified graphene can be incorporated into the semiconductor industry in the future.

Besides the material's application perspective, the electronic structure of graphene has attracted a lot of attention from the condensed matter community for many reasons[5]. The relation between energy and momentum in graphene is very different from any other material due to the honeycomb lattice structure. For non-relativistic electrons moving in free space the energy E is related to the momentum p by the classical relation: $E = p^2/(2m)$, where m is the electron mass. This relation is extremely robust and in many materials is obeyed even in the presence of interactions between electrons and lattice (ions) and among electrons themselves. This classical relation persists, however, with a slight modification: the electron mass m is replaced by an effective mass $m^*$ which reflects the change in the inertia of the electron due to the presence of an environment. In the honeycomb lattice this energy-momentum relation changes to something completely unexpected (see Fig. 4): $E = \pm v |p|$ (the plus and minus sign refers to the two cones or bands of graphene), where v is the so called Fermi-Dirac velocity that depends on the material properties. This relation is the same obeyed by massless relativistic particles (such as the neutrino) with the speed of light c replaced by v (in graphene, $v \sim c/300$). This change in the energy-momentum relation has profound consequences for the physics of graphene electrons.

The first obvious difference between normal metals and graphene is the fact that while metals usually require just one energy band to describe them (see Fig. 4), graphene, like a relativistic system has two bands: one of particles and another or anti-particles (which in solid state physics are called holes). In neutral graphene, the particle band is empty while the hole band is full. Graphene is also different from ordinary semiconductors which usually also require two bands for their description (conduction and valence bands) because it does not have a gap in the spectrum. Hence, graphene is a hybrid between a metal and a semiconductor, and many of its unusual properties derive from this fact. Because graphene does not have a gap in the spectrum, for many practical purposes it works as a metal. However, for device applications, where large on-and-off current ratios are required, this is a drawback.

A lot of the research effort in the graphene field has been focused in trying to generate a gap in the graphene band structure so that it energy-momentum becomes $E = \pm \sqrt{v^2 p^2 + \Delta^2}$, where the gap between the two bands is $2\Delta$. At small momentum, i.e. $p \ll \Delta/v$, this energy-momentum relation mimics an ordinary semiconductor (see Fig. 4), $E \simeq \pm \Delta \pm p^2/(2m^*)$ where $m^* = \Delta/v^2$, the analogue of Einstein's energy-mass relation. At larger momentum, $p \gg \Delta/v$, however, the gapped graphene recovers its linear energy-momentum relation, in contrast with the usual semiconductor. Hence, gapped graphene is not exactly an ordinary semiconductor (see Fig.4).

There are a few ways to generate a gap in graphene. Conceptually, the simplest one has to do with the fact that the honeycomb lattice is made out of two identical interpenetrating triangular sub-lattices (see Fig.5). If these two sub-lattices are not identical a gap can open in the spectrum. A possible way to produce this effect is to chose a specific substrate that generates an electrostatic potential that is different in different sub-lattices (so that the sub-lattice symmetry is broken) and a gap opens in the spectrum.

It is also worth pointing out that the unusual energy-momentum in graphene has also important consequences for the interactions between electrons. Consider that the graphene lattice is doped either with electrons or holes with a concentration, $\sigma = 1/l^2$, of electrons per unit of area ($l$ is the average distance between electrons). In quantum mechanics the momentum of a particle is related to its wavelength $\lambda$ by $p = \hbar/\lambda$, where $\hbar$ is Planck's constant. In a normal metal or semiconductor, the energy-momentum relation then implies that $E_M = \hbar^2/(2m^*\lambda^2)$. The Coulomb interaction between two electrons separated by a distance r behaves as $U = e^2/(\varepsilon r)$, where e is the electric charge and $\varepsilon$ is the dielectric constant of the medium. Notice that for electrons separated by a distance $l$ the average Coulomb energy scales like $E_C \propto e^2/(\varepsilon l) = e^2 \sigma^{1/2}/\varepsilon$. Hence, for a normal metal or semiconductor where $\lambda \simeq l$, the kinetic energy scales as $K_M \propto \hbar^2/(m^* l^2) = \hbar^2 \sigma/m^*$. Thus, the relative strength of the interactions is determined by the electronic density. Strong electron-electron interactions occur when the Coulomb dominates the kinetic energy, $E_C \gg K_M$, that is, at low densities, $\sigma \ll \sigma_0 = [m^* e^2/(\varepsilon \hbar^2)]^{1/2}$. This low density region of the two-dimensional electron gas can have a series of different phases ranging from Wigner crystals and charge density waves, to magnetism[13]. At higher densities, $\sigma \gg \sigma_0$, the two dimensional electron gas behaves as a system of weakly interacting particles, the Fermi liquid.

In graphene the situation is rather different. The relation between energy and wavelength is $E_G = v \hbar/\lambda$ and therefore the average kinetic energy is given by $K_G \propto v \hbar/l = v \hbar \sigma^{1/2}$. Since the Coulomb energy does not care whether the system is graphene or something else, the ratio between Coulomb and kinetic energy is independent of the density and given by the so-called graphene's fine structure constant[5]: $g = E_G/K_G = e^2/(\varepsilon \hbar v)$. Notice that in this case, the cases of strong and weak interactions is not determined by the density, but by the value of the dielectric function. Therefore, the nature of the electronic states in graphene is rather dependent on the nature of the environment where graphene is laid. For graphene on top of $SiO_2$, where $\varepsilon \cong 3$, we would have $g_{SiO} \cong 0.7$ and hence we can conclude that the Coulomb interactions are relatively weak, while for a suspended graphene sample with $\varepsilon \cong 1$ we have $g_0 \cong 2.1$ and the Coulomb interactions are clearly outside the perturbative regime[14]. Therefore, unlike the two-dimensional electron gas, the substrate can play a fundamental role on the nature of the electronic states in graphene. A similar issue occurs in graphene in the presence of a perpendicular magnetic field B. Just like in the two-dimensional electron gas, the presence of a magnetic field is singular and leads to the immediate creation quantized energy levels, the so-called Landau levels, whose energy is given by : $\hbar \omega_C = \sqrt{2} v/l_C$ where $l_C = [c/(e B)]^{1/2}$ is the cyclotron length. The role played by $l$ in the case of B=0 is now played by $l_C$ and the question of the nature of the electronic ground state is one of the most challenging problems in graphene research.

One of the unique aspects of graphene is its membrane-like nature. Being only one atom thick, distortions of the graphene lattice out of the plane cost very little energy[15]. Hence, graphene is probably the only example of a metallic membrane[9] (most biological membranes are highly insulating). In the presence of wrinkles and ripples the electronic structure changes because the electronic[16] orbitals become distorted as well[9]. Hence, unlike any other solid state material, the electronic properties of graphene are dependent on its conformation. On the one hand, this effect can be detrimental to electronic motion because electrons can be scattered by local uncontrolled distortions. This would happen when graphene is laid on top of a rough substrate[17] (as is the case of $SiO_2$) or if graphene is under random shear strain[18] (as is the case of suspended samples). On the other hand, the coupling between structure, strain and electronic properties can also be put to good use in what is called "strain engineering"[19], that is, gaps in the electronic spectrum can be produced by either sufficiently large uniform strain[20] (that may, unfortunately, also rip graphene apart) or by small but non-uniform strain[21]. Strain, even if it does not generate gaps, can also induce strong anisotropies in the charge transport that can be used for several applications[22]. In Fig. 6 we show a simple device based on strain engineering.

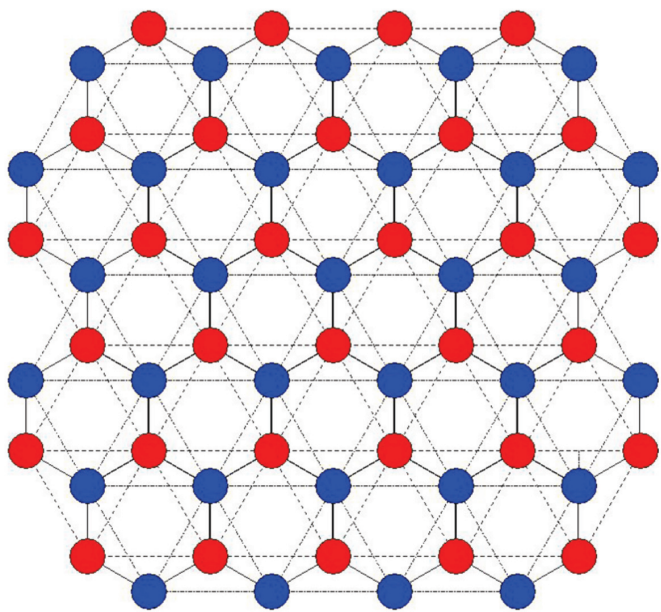

Fig. 5 Honeycomb lattice. Red and blue circles represent the two different sub-lattices. The dashed lines, in the form of David's stars, clearly show the two interpenetrating triangular sub-lattices.

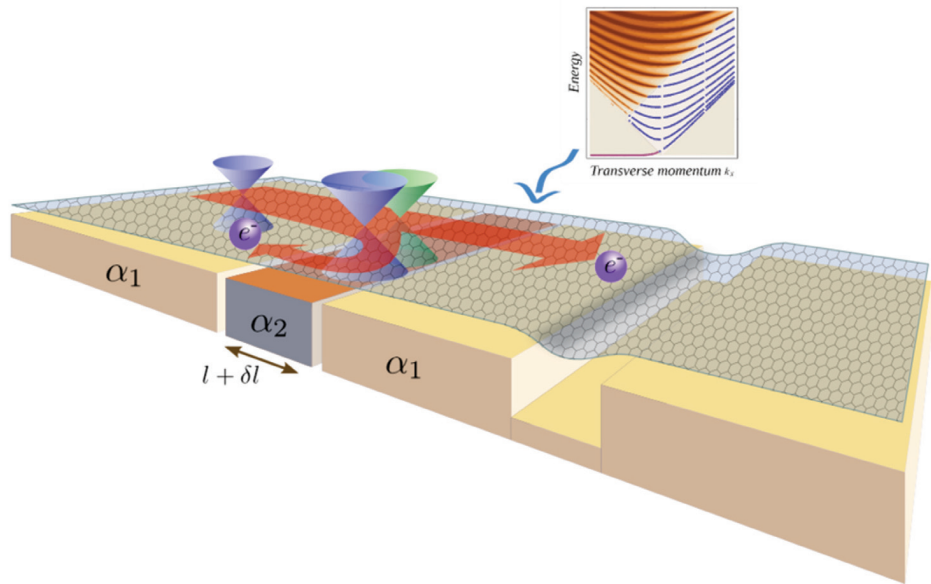

Fig. 6. Depiction of a a possible device that induces local strain. A substrate is fabricated out of two materials with different thermal expansion coefficients ($\alpha 1$, $\alpha 2$), or containing selected trenches. Any change in temperature will cause a different amount of linear expansion, such that the two regions of graphene above are deformed by different amounts. This generates local strain, which will scatter the Dirac electrons propagating across that region. This creates a transport gap in the conductance at low densities. At the same time, for a range of incoming velocities, the electrons can become confined, just exactly as in a graphene nanoribbon with an electronic spectrum as shown in the inset[19].

Graphene research is one of the fastest growing areas in science, but it is still a young field. There are many challenges and opportunities for investigation, because graphene is not a standard solid state material. Electrons in graphene do not behave in the same way as in ordinary metals and semiconductors, because of the unusual energy-momentum relation. From this perspective, the theory of metals has to be rewritten for it. Graphene is a metallic membrane and its soft nature affects directly its electronic properties. The literature on metallic membranes is incipient. Because of our ignorance on the basic nature and the limitations of this material, there is a lot of hoopla in the media. A myriad of applications, many of them pure fantasy, have been spread all over the internet as a quick search would reveal. For these, and many other reasons, the graphene field has been surrounded by a lot of hype but also hope. Hope that this material is unveiling the dawn of a new era where carbon, the element of life, also becomes an element of progress.

## Acknowledgements

*I would like to thank J. Nilsson for Fig. 3, V. Pereira for Fig.6, and E. Mucciolo for bringing to my attention Ref. 1. This research was partially supported by the U.S. Department of Energy under the grant DE-FG02-08ER46512 and ONR grant MURI N00014-09-1-1063.*